\documentclass[preprint,pre,aps,superscriptaddress,a4paper]{revtex4}

\usepackage[utf8x]{inputenc}
\usepackage{graphicx, subfigure}
\usepackage{epstopdf}
\usepackage{amsmath}
\usepackage{amsfonts}
\usepackage{color}
\newcommand{\ro}{\hat{\rho}}
\begin{document}
\title{Effects of rigid or adaptive confinement on colloidal self-assembly.
Fixed vs. fluctuating number of confined particles}
\author{J. P\c ekalski}
\affiliation{Institute of Physical Chemistry,
 Polish Academy of Sciences, 01-224 Warszawa, Poland}
 \author{N. G. Almarza}
\affiliation{Instituto de Qu{\'\i}mica F{\'\i}sica Rocasolano, CSIC, Serrano 119, E-28006 Madrid, Spain }
\author{ A. Ciach}
\affiliation{Institute of Physical Chemistry,
 Polish Academy of Sciences, 01-224 Warszawa, Poland}
\date{\today}
\begin{abstract}
The effects of confinement on colloidal self-assembly in the case of fixed  number of confined particles
are studied in the one dimensional lattice model  solved exactly in the Grand Canonical Ensemble (GCE) in 
[J. P{\c e}kalski et al. J. Chem. Phys. \textbf{142}, 014903 (2015)]. 
The model considers a pair interaction defined by a short-range attraction plus a longer-range repulsion.
 We consider thermodynamic states corresponding to self-assembly into clusters. Both, fixed and adaptive boundaries are studied. 
For fixed boundaries, there are particular states in which, for equal average densities, 
the number of clusters in the GCE is larger than
in the Canonical Ensemble.
The dependence of pressure on density has a different form when the system size changes with fixed number of particles
and when the number of particles changes with fixed size of the system.
In the former case the pressure has a nonmonotonic dependence on the system size.
The anomalous increase of pressure for expanding system is accompanied by formation of a larger number of smaller clusters.
In the case of elastic confining surfaces we observe a bistability, i.e. two significantly different
system sizes occur with almost 
the same probability. 
The mechanism of the bistability in the closed system is different to that of the  case of permeable walls, where 
the two equilibrium system sizes correspond to a 
different number of particles.
\end{abstract}
\maketitle
\section{Introduction}

Confinement has a significant effect on fluids when the separation between the confining surfaces is 
comparable with the characteristic
structural length
of the confined fluid ~\cite{evans:90:0,israel:11:0}. 
In simple fluids such pore thicknesses are of order of {\AA}ngstr\"oms.  The effects of confinement depend on whether the  boundaries are
rigid or adaptive. In the latter case
mechanical equilibrium between the stress in the solid walls resulting from the swelling or shrinking of the pore, and
 the solvation force~\cite{evans:90:0,israel:11:0} induced on the walls by the confined fluid can lead to contraction or expansion of
 the pore~\cite{kowalczyk:08:0}. 
  Similar dependence on the type of confinement is expected in the case of self-assembling systems, but on different length and energy scales
  ~\cite{kekicheff:89:0,kekicheff:95:1,tasinkevych:99:0,ciach:04:3,archer:08:0,imperio:07:0,perkin:13:0}.
  The packing effects of molecules are replaced in this case by the packing
  effects of micelles, clusters or layers that are much larger and softer. 
  
 In this work we focus on colloidal self-assembly in thermodynamic conditions corresponding to 
 self-assembly into clusters or layers in the bulk\cite{campbell:05:0,stradner:04:0,imperio:04:0,archer:07:0,archer:08:0,ciach:10:1}. 
We study rigid and elastic boundaries, and  
  consider separations of the confining walls comparable with a few structural units.
 The adaptive confinement in this case means confining surfaces that are soft and
separated by hundreds of nanometers. Important examples of such a confinement exist in organella or  in lipid or polymeric vesicles. 
 One can expect that the shape of the outer membrane and the ordering inside the vesicle or organellum influence each other  in a 
 way that depends on the elasticity of the membrane. We shall consider self-assembly in systems with fixed boundaries and with boundaries with
 different elasticity, from stiff to very soft.
 
 Some of the membranes
 are permeable, while some other are not. Thus, a question arises if fluctuations of the number of confined particles have any effect on
 the properties of a self-assembling system confined by    rigid or adaptive boundaries. This question motivates our present study.
 We ask how the effects of confinement 
 on the self-assembling system depend on the contact with a reservoir of particles. 
 
 The above problems are very difficult for realistic models of self-assembling systems confined by elastic boundaries.
 In order to gain some first insight, however, one can consider simplified models. In this work we consider the simple generic model
 of the system with competing short-range attraction and longer-range repulsion (SALR) that can be solved exactly \cite{pekalski:13:0}. 
 The model was solved exactly and systematically analyzed in the bulk \cite{pekalski:13:0,pekalski:14:0,almarza:14:0}
 and in confinement~\cite{pekalski:15:0} in the grand
 canonical ensemble (GCE). 
Recently a procedure to extract full Canonical information from Grand-Canonical results
has been proposed in Ref.~\cite{heras:14:0}.
In principle, it is possible to apply this technique to our
exact GCE results. However, in the current problem the presence of energetic terms and low temperature pose
numerical difficulties in the mapping procedure between the results in both ensembles. For this reason we decide to perform
MC simulations in the canonical ensemble (CE).

 In the case of fixed boundaries we address the question of how the fluctuations of the positions and sizes of the clusters (their dynamical 
 assembly and dissociation) are coupled with the fluctuations of the number of particles in the system.
  In the case of inhomogeneous distribution of particles it is not a priori obvious that the largest fluctuations in the 
 total number of particles lead to the largest differences between the density profiles in the canonical and grand 
 canonical ensembles.
 We shall compare the  density profiles
 and the pressure in the GCE and CE ensembles, with
 the average number of particles
 in the GCE equal to the number of particles $N_0$ in the CE. We shall pay particular attention to values of $N_0$ that are 
 too small or too large for a given system size $L$ for formation of periodically distributed layers of particles that
 minimizes the system energy. Roughly speaking, in the SALR systems the minimum energy is assumed when  the individual
 clusters have the lowest energy  (no intra-cluster repulsion), 
 and are separated by the smallest distances corresponding to no inter-cluster repulsion. 
 When the number of particles
 is too small or too large for a given $L$ for formation of the optimal bulk structure, 
 some structural deformations must occur. 
 Our purpose is to compare these deformations and their effect on the mechanical properties in the GCE and CE.
 
 Adaptive boundaries  were studied in Ref.[\onlinecite{williams:13:0}] in the case of fixed number of  discs surrounded by particles kept 
 by  laser tweezers. The system exhibited bistability; either hexagonal structure (and modified boundaries)
 or concentric rings of particles occurred. We observed 
 a different bistability in Ref. \cite{pekalski:15:0}, where we studied  particles interacting with the SALR potential and
 confined by
 boundaries, whose separation could be varied at the cost of elastic energy.  When the average number of particles
 is too small or too large for  given $L$ for formation of the optimal bulk structure, 
 some compromise must occur between the excess of the free energy associated with the structural deformations,
 and the elastic energy cost due to adjusting the size of the compartment to the optimal structure of the confined fluid.
We obtained exact results for a one dimensional (1d) model in GCE, and required mechanical equilibrium between the fluid-induced solvation force 
and the elastic force present when the system shrinks or expands.  We found that
 when the equilibrium width of the empty container corresponded to the largest stress of the confined colloidal system,
  two significantly different system sizes were almost equally probable. 
 The large size fluctuations are accompanied by absorption or 
 evaporation of a whole cluster.
 Clearly, for fixed number of particles such large size fluctuations are not possible. Hence, the permeability  
 plays an important role in the case of elastic boundaries. In this work we verify if in the case  of fixed number of particles 
 the  bistability of the system size can still exist due to some other mechanism.
 
  
  We present the model and the simulation methods in Sec. \ref{sec2}.
 The density profiles are described and compared
 with the results obtained in the GCE in Sec. \ref{sec3}. The mechanical properties are discussed in Sec. \ref{sec4}.
 We  compare the dependence of pressure on density in the CE (fixed $N$ and varying $L$) and in the GCE for several fixed values of $L$
 and varying $\langle N\rangle$.  
 The dependence of pressure on the system size
 for fixed $N$ or $\langle N\rangle$ in the CE or GCE respectively is also discussed. 
 In Sec. \ref{sec5} we consider elastic boundaries and compute the average system size as a function of $N$ 
 for various elastic constants of the walls.
 For selected cases the histograms for the fluctuating width of the system are presented. 
Sec. \ref{sec6} contains a short summary and concluding remarks.

\section{The model and the methods} \label{sec2}
In this section we briefly describe the model and the methods used for its analysis. 
More detailed descriptions of the model and the transfer matrix method used for finding 
the exact solutions of the model in the GCE can be found in \cite{pekalski:15:0}.
\subsection{The model}
We consider  a one-dimensional (1d) lattice model and assume an isotropic effective 
pair interaction potential of the SALR type. 
The ranges of the competing attractive and repulsive parts are chosen such that small clusters are 
formed for some range of thermodynamic variables. 
Namely, we assume that the nearest neighbors attract each other and the third neighbors repel each other, i.e.
the effective pair interaction potential is
\begin{equation}
\label{V}
V(\Delta{ x}) = \left\{ \begin{array}{ll}
-J_1 & \textrm{for $|\Delta{ x}| = 1$},\\
+J_2 & \textrm{for $|\Delta{ x}| = 3$},\\
0 & \textrm{otherwise,}
\end{array} \right.
\end{equation}
where $J_1$ is the energy of attraction, $J_2$ is the energy of repulsion and the unit distance is 
the particle diameter $\sigma$, and $\Delta{ x}$ is the distance between two sites of the lattice.
Such a shape of the effective potential can be found for charged particles in solvents that induce 
short range attraction between the particles,
e.g. for lysozyme molecules in water \cite{shukla:08:0} or for colloids in a solvent containing short-chain 
polymers~\cite{stradner:04:0}.

In Ref.\cite {pekalski:13:0, pekalski:14:0, almarza:14:0}  we assumed periodic boundary conditions and
extensively investigated the spontaneous pattern formation of particles in the bulk.
In order to study the effects of confinement on the self-assembled structures, 
we changed the boundary conditions of the model from periodic to rigid or elastic 
in Ref.\cite{pekalski:15:0}. 
In the current study we continue the investigation of the confined system with particles interacting \textit{via} 
the pair potential given by Eq. (\ref{V}). 
The confinement is assumed to be electrically neutral, 
hence the interaction of the particles with the walls is short range. 
The Hamiltonians in the canonical, and Grand canonical ensembles take the forms:
\begin{equation}
  U [\{\hat\rho\}]  = 
\frac{1}{2} \sum_{x=1}^L\sum_{x'=1}^L \hat{\rho}( x) V ( x-x')\hat{\rho}(x') + h_1 \hat{\rho}(1) + h_L \hat{\rho}(L)  
\hspace{1cm}  ({\textrm fixed}\; N); 
\label{Hcan}
\end{equation}
\begin{equation}
   H [\{\hat\rho\}]  = 
   U [\{\hat\rho\}]  - \mu \sum_{x=1}^L \hat\rho (x). 
 \hspace{1cm}({\textrm fixed}\; \mu ),
\label{H}
  \end{equation}
where $h_1$ and $h_L$ are the energies of the interactions between the confining walls and the  particles 
located at the $1$-st and the $L$-th site respectively, $\ro(x)$ is the occupancy operator such 
that $\ro(x)=1$ if the $x$-th site is occupied and $\ro(x)=0$ otherwise.
The number of particles $N$ (fixed in the canonical ensemble) is given by:
\begin{equation}
N = \sum_{x=1}^L \hat\rho (x).
\end{equation}

As in Ref. [\onlinecite{pekalski:15:0}], we assume that the confinement can be either rigid or elastic.
For rigid boundary conditions the distance $L$ between the confining walls is fixed, while in the case of elastic walls 
we assume that it may oscillate around $L_0$, which is the equilibrium width of an empty system. 
The Hamiltonian of the system with elastic boundary conditions is for the case of permeable walls: 
\begin{equation}
\mathcal{H} = H+k(L-L_0)^2,
\label{elasticH}
\end{equation}
where $k$ is the elastic constant, and for the fixed number of particles the Hamiltonian takes the form:
\begin{equation}
\mathcal{U} = U+k(L-L_0)^2.
\label{elasticU}
\end{equation}
 
We set the energy of attraction $J_1$ as energy unit and introduce the following dimensionless variables 
\begin{eqnarray}
\label{dimensionless}
T^*&=&k_BT/J_1,\quad \beta^*=J_1/k_BT, \quad J^*=J_2/J_1, \quad \\
h_1^*&=&h_1/J_1, \quad  h_L^*=h_L/J_1,\quad \mu^*=\mu/J_1
 \end{eqnarray}
where $k_B$ is the Boltzmann's constant and $T$ is the temperature. From now on we  set $J^*=3$, and $h_1^*=h^*_L = \pm 1$, and study the role of  the temperature and the chemical potential
 or the number of particles.
For the selected parameters the energy (\ref{Hcan}) takes the minimum when clusters composed of 3 particles are 
separated by 3 empty sites, 
and a cluster is attached to each boundary. Such an optimal structure is possible only for $L=2N-3$.

 \subsection{The computational method}

In order to find the exact solution of the model in the GCE for $T>0$ the transfer matrix method was applied.
We have found exact expressions for the partition function and the local density. The details of the derivations
are provided in the the appendix  of Ref. [\onlinecite{pekalski:15:0}]. 
Here we will present only the final formula for the partition function with
a brief description of the notation.
The partition function for the system of size $L=3M+j$, with $M$ a natural number and $j=0,1,2$ is given by
\begin{eqnarray}
\Xi =\! \sum_{\hat S(1)}  \sum_{\hat S(M)}'  e^{\beta^* \ro(1) h^*_{1} } {\bf T}^{M-1}[\hat S(1),\hat S(M)] 
e^{\beta^* \ro(L) h_L^*}e^{\beta^* H^*_j[\hat S(M)]},
\label{ss}
\end{eqnarray}
where $\hat S(n)=(\hat\rho(3n-2),\hat\rho(3n-1),\hat\rho(3n))$ with $n = 1,\ldots M$ and  ${\bf{T}}$ is the 
transfer matrix with the matrix elements defined as
\begin{eqnarray}
\label{T}
{\bf T}(\hat S(n),\hat S(n+1))\equiv e^{-\beta^* H^*_t[\hat S(n),\hat S(n+1)]},
\end{eqnarray}
where
 \begin{eqnarray}
 H^*_t[\hat S(n),\hat S(n+1)]=\sum_{x=3n-2}^{3n}\big[
-\hat\rho(x)\hat\rho(x+1)+J^*\hat\rho(x)\hat\rho(x+3)-\mu^*\hat\rho(x)
\big].
\end{eqnarray}
$\sum_{\hat S(M)}^{'}$ and $H_j^*$ depend on $j$ and the rather lengthy expressions are provided in Appendix \ref{app1}.
Having a formula for the partition function one can derive exact expressions for the pressure and the local density. 
For more details as well as for asymptotic expressions of the exact solutions see \cite{pekalski:15:0}.

 \subsection{The simulation methods}
 In the current study we compare the exact solution obtained in GCE with results
of MC simulation in CE.
In addition to the simulations in the CE we also carried out some runs in the GCE with the
aim of cross-checking the consistency between the simulation codes and the numerical treatments
based on the transfer matrix methods.
The simulation
procedures make use of  the Metropolis criterion
\cite{metropolis:53:0} implemented for two kinds of MC steps: translations
of one  particle in the Canonical Ensemble and particle insertions or deletions
in the Grand Canonical ensemble.
 For each step a trial configuration is generated and it is accepted with probability:
$\min\left[ 1,\exp(-\beta \Delta H) \right]$, where $\Delta H$ is the change of the Hamiltonian in
the trial step.
The trial configuration for a particle \textit{translation} is generated
by moving a randomly chosen particle to a randomly chosen empty site, which is equivalent
to choose two sites of the lattice $x$ and $x'$ so that $\hat{\rho}(x) \ne \hat{\rho}(x')$,
and interchange their occupancy states, so that $\hat{\rho}^{\text trial}(x) =
\hat{\rho}(x')$, and $\hat{\rho}^{\text trial}(x')=\hat{\rho}(x)$.
A Monte Carlo step in the Grand Canonical simulations implies the
insertion or deletion of one particle, this is achieved by choosing at random
one site of the lattice, $x$,  and generating the trial configuration
by \textit{flipping} its
occupancy state from its current value $\hat{\rho}(x)$ to the trial value
$\hat{\rho}^{\text trial}(x)=1-\hat{\rho}(x)$.

The computation of pressure from simulation of lattice models is usually carried
out by means of the integration of the Grand Potential in the GCE, because the known
relation for the canonical ensemble:
$\beta p = - \left( \partial (\beta A)/\partial V\right)_{N,T}$
(where $A$ is the Helmholtz free energy, and $V$ the \textit{generalized} volume)
is hard to translate into an efficient numerical procedure due to the discreteness of 
the volume in lattice systems.
We have found, however, that for our 1d system it is feasible
to compute the pressure $p(N,L,T)$ in the CE, by an algorithm based on the
discretization of the derivative of $A(N,L,T)$ with respect
to the system size as:
\begin{equation}
\label{p}
 p_{\pm}(N,L,T) = \mp [A(N,L\pm1,T) - A(N,L,T)].
\end{equation}
The pressure, in the terms of the canonical partition function $Q(N,L,T)$ can be written as
\begin{equation}
p_{\pm}(N,L,T) = \pm \frac{1}{\beta} \ln \frac{Q(N,L\pm1,T)}{Q(N,L,T)}.
\label{press}
\end{equation}
The two ways of discretization, $p_+$ and $p_-$, lead to two different methods of computing
the pressure, the \textit{virtual expansion} and the \textit{virtual contraction} respectively,
from a direct analysis of the configurations from a simulation run at conditions $(N,L,T)$.

In the virtual expansion an empty site is added at a randomly chosen position
of the system.  For a confined system of size $L$, there are $L+1$ possibilities of performing
such an insertion, namely $L-1$ cases where the inserted site is located between two sites
of the system plus two insertions between the walls and the first or last site.
Considering the $L+1$ possible ways of inserting an empty site on each of the
microstates of the system of size $L$ with $N$ occupied sites
(with $0\le N \le L$), we get $L+1-N$ identical copies of each of the the microstates
of the system of size $L+1$ and $N$ occupied sites (see Appendix \ref{app2}).
Let us denote by $\vec{\rho}_L$ a given configuration of the system with $N$ particles
and $L$ sites, with potential energy given by $U(\vec{\rho}_L)$.
If we define
$\vec{\rho}_{L+1}( \vec{\rho}_L, k)$  as the configuration with $N$ particles of a system with
$L+1$ sites built from $\vec{\rho}_L$ by inserting a site at position $k$,
and denote by
$\sum_{\vec{\rho}_L}$ the sum over all possible microstates of this system with $N$ particles
and $L$ sites,
then we can write Eq. (\ref{press}) as:
\begin{eqnarray}
p_{+}(N,L,T) &=& \frac{1}{\beta} \ln 
\frac{\sum_{ \vec{\rho}_L} \exp\left[-\beta  U(  \vec{\rho}_L)\right] 
\sum_{k=0}^{L} \exp \left[-\beta  U( \vec{\rho}_{L+1}(\vec{\rho}_L,k)) 
 +\beta   U(\vec{\rho}_L)\right]}
{(L+1-N)\sum_{\vec{\rho}_L}\exp\left[-\beta  U(\vec{\rho}_L)\right]} \\
&=& k_B T \ln  \left\langle \frac{L+1 }{L+1-N} 
 \exp \left[ -\beta \Delta  U_{ins}\right] \right\rangle_L, 
\end{eqnarray}
where $\langle \cdot \rangle_L$ is the average value of $\cdot$ when sampled on a system of size $L$,
and $ \Delta U_{ins}$ is the difference between the energies of the systems with $L+1$ and $L$ sites.
Analogously, a formula for pressure $p_-$ computed by the \textit{virtual contraction}  scheme can be derived
\begin{equation}
p_-(N,L,T) = -k_B T \ln \left\langle \frac{L-N}{L} \exp \left[ -\beta \Delta U_{del} \right]\right \rangle_L,
\end{equation}
where $\Delta  U_{del} =  U(\vec{\rho}_{L-1})- U(\vec{\rho}_L)$ is the variation of energy when
a configuration $\vec{\rho}_{L-1}$,  of $N$ particles and $L-1$ sites is generated by eliminating one of the 
empty sites from the configuration $\vec{\rho}_L$ of a system with $N$ particles and $L$ sites.
The virtual contraction method 
is inefficient at high densities, therefore we used it only for verification of the results 
obtained \textit{via} the virtual expansion method, since by construction we expect:
\begin{equation}
p_+(N,L.T) = p_-(N,L+1,T).
\end{equation}
In order to calculate properties of the system with elastic boundary conditions described 
by the Hamiltonian $\mathcal{U}$ given in Eq. (\ref{elasticU}), 
one needs to perform two additional types of MC steps. 
The first one is the \textit{intercalation} of an empty site into a randomly chosen place of the system,
the second one is the \textit{removal} of a randomly chosen empty site of the system.
The acceptance probability of the first move for system with $L$ sites and $N$ particles is
\begin{equation}
\mathcal{A}(L+1|L) = \textrm{min} \Big\{ 1,\exp [-\beta \Delta \mathcal{U}] \frac{L+1}{L+1-N} \Big\},
\end{equation}
where $\Delta \mathcal{U}$ is the change of the energy after the size modification.
Analogously the probability of acceptance of the move in which an empty site is removed is given by:
\begin{equation}
\mathcal{A}(L-1 | L) = \textrm{min} \Big\{1,\exp [-\beta \Delta \mathcal{U}] \frac{L-N}{L}\Big\}.
\end{equation}
Notice that this procedure resembles a lattice version of isothermal-isobaric $(NpT)$ simulation, in which the energy term
introduced through the elastic force plays the role of the external field. 

\section{Distribution of particles between rigid walls} \label{sec3}
In this section we consider a system containing  $N_0$ particles  between rigid walls separated by a fixed distance $L$.
 The main question is how the particles self-assemble if $N_0$ is such that the equilibrium bulk structure is not possible.
 The distribution of particles for fixed $N_0$ will be compared with the distribution of particles in the open system,
 where the number of particles $N$ fluctuates in such a way that $\langle N\rangle=N_0$. 

\begin{figure}[ht]
    \begin{center}
      \includegraphics[scale=1]{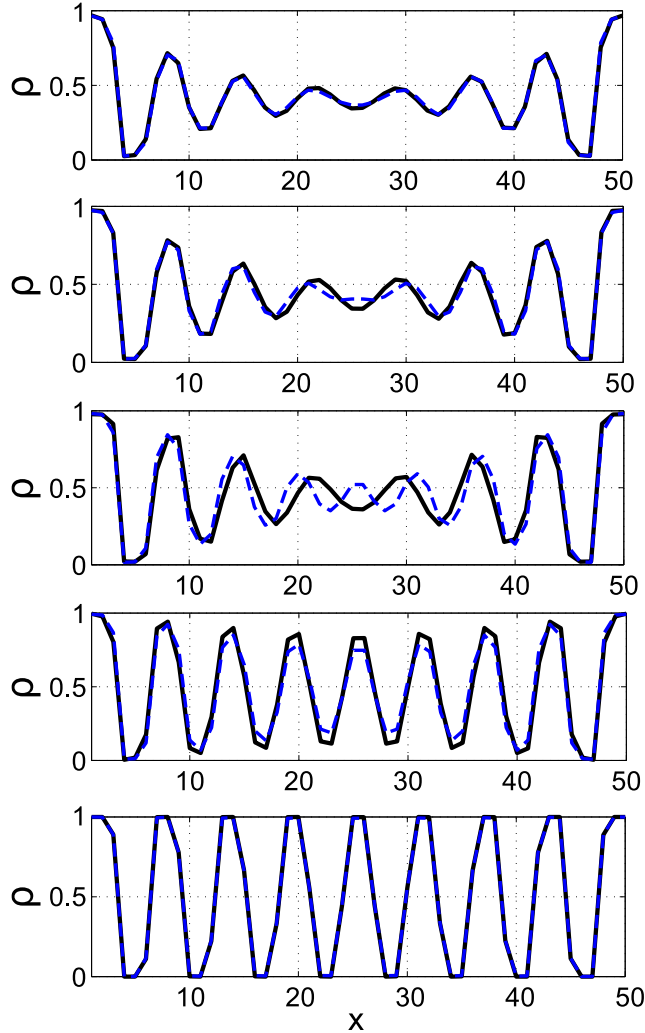}
      \caption{ Comparison of the GCE (dashed line) and the CE (solid line) density profiles for 
       $L = 50$ at  $T^*=0.3$ and  $N=22$ or $\mu^*=-0.33904$ (a), $N=23$ or $\mu^*=-0.21127$ (b), $N=24$ or $\mu^*=-0.02632$ 
      (c), $N=25$ or $\mu^*=0.22159$ (d), $N=26$  or $\mu^*=1.08857$(e).
Repulsion to attraction ratio $J^*=3$ and attractive walls with $h_1^*=h_L^*=-1$ are considered for all the cases. }
      \label{ensemble_compar2a}
    \end{center}
  \end{figure}

\begin{figure}[ht]
    \begin{center}
      \includegraphics[scale=1]{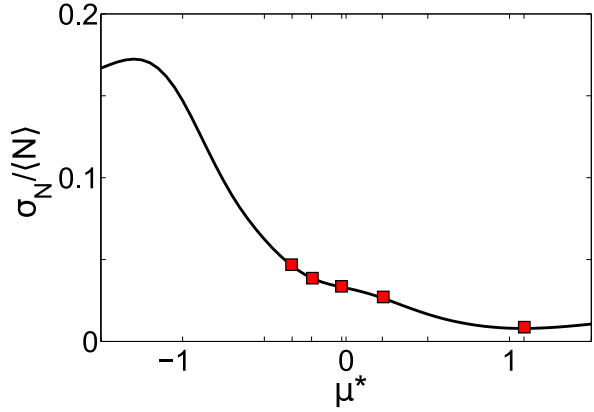}
      \caption{The standard deviation of the number of particles $\sigma_N$ divided by the average number of particles 
      $\langle N \rangle$ for $L=50$, $T^*=0.3$, $J^*=3$ and attractive walls with $h_1^*=h_L^*=-1$.
      The red squares indicate the values of the chemical
      potential taken for the density profiles in Fig. \ref{ensemble_compar2a}.}
      \label{ensemble_compar3b}
    \end{center}
  \end{figure}

 In Fig.\ref{ensemble_compar2a} the density profiles obtained by the MC simulations in the CE are compared 
with the exact results obtained in the GCE by the transfer matrix method described in Ref.\cite{pekalski:15:0}. 
We chose $L=50$, $T^*=0.3$ and several values of $N_0$. 
In each case the
chemical potential in the GCE was fixed to the value that corresponds to $\langle N\rangle=N_0$.
 We used the exact expression for density as a function of the chemical potential that was obtained 
in Ref.[\onlinecite{pekalski:15:0}]. 
For $L=51$ the optimal number of clusters for the considered range of $\mu^*$ is $9$ (hence $N=27$), 
since the sequence of three occupied sites followed by three empty sites can be formed,
with two clusters adsorbed at the attractive surfaces. 
The energy for such a structure assumes a minimum (there are as many attracting pairs as possible with no repulsion). 
For $L=50$ only a small defect in the ordered structure occurs. 
We can see a very good agreement between the two ensembles for small
as well as for large number of particles. 
For $N=24$ corresponding to eight  clusters, however, the number of maxima in the GCE is larger than in the CE.
This result is even more surprising when we consider the fluctuation of the number of particles in the
GCE (Fig.\ref{ensemble_compar3b}). 
One can see that the largest discrepancy between the density profiles does not occur for the 
largest fluctuation of the number of confined particles in the GCE.
        \begin{figure}[ht]
    \begin{center}
      \includegraphics[scale=1]{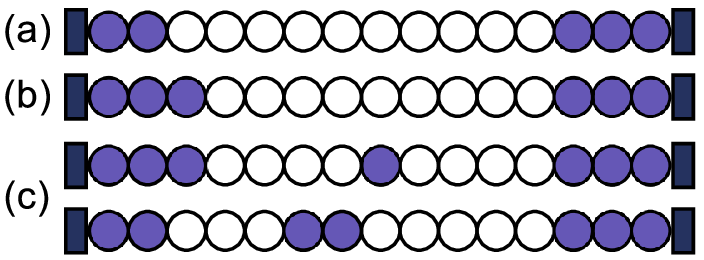}
      \caption{Typical microscopic states present in the GS of the model with $J^*=3$,
      attractive walls with $h_1^*=h_L^*=-1$ and $L=15$ in the CE with  
  $N=5$ (a), $N =6$ (b) and $N=7$ (c). For $N=5,7$ the GS is degenerate. }
      \label{GS15}
    \end{center}
  \end{figure} 
 
   In order to understand why the distributions of the particles in the two ensembles are different when one cluster in the CE 
 ``is missing'',
 let us consider the ground state (GS), $T^*=0$. 
 The microstates present in the GS for $L=15$ and 
 $N=5,6,7$ are shown in Fig.\ref{GS15}. The GS in Fig.\ref{GS15} shows that even small fluctuations of the number of particles - addition of one particle in our case 
  - can lead
  to a significant change in the distribution of the particles. This is the case when $\langle N\rangle$ is a multiple of $3$,
  and there is a free space for an extra cluster (with no cluster-cluster repulsion).
  When one additional particle enters the system, the energy change is  $J^*-1$ when one of the clusters grows to contain 4 particles, 
   or  $0$, when the new particle is sufficiently far from the clusters, or one of the clusters together 
   with the new particle form two clusters composed of two particles.
  In Fig.\ref{GS15}, bottom row, the two latter cases are shown. The states shown in Fig.\ref{GS15}
  are energetically favorable for $J^*>1$. Thus, in the GCE with $\langle N\rangle=6$ such microstates will appear quite often.
  As a result, an additional maximum in the average density
  profile occurs.
  
  The above simple considerations show that spatial distribution of particles in the CE and GCE  can be qualitatively different.
  This qualitative difference is not present for the largest fluctuation of the number of particles in the GCE. Even a small 
  fluctuation of the number of particles can lead to a change of the number of clusters, because
  the sizes of the clusters can fluctuate, especially when $N$ is not a multiple of 3. When $\langle N\rangle$ is  small enough,
  the additional
  clusters can occupy the empty space and no inter-cluster repulsion appears.

\section{Equation of state in a system confined by rigid  walls} \label{sec4}

In this section we compute pressure for fixed number of confined particles $N_0$ as a function of the distance between the
confining surfaces $L$. From these results we obtain the pressure
 as a function of density, $p(\rho)$, for given $N_0$. For comparison
we present $p(\rho)$ calculated exactly in the GCE by the transfer matrix method described in Ref.\cite{pekalski:15:0}.
In the GCE we consider fixed $L$ and $\mu$, and calculate $p(\mu)$ and $\rho(\mu)$ to obtain $p(\rho)$ for given $L$. 
The shape of  $p(\rho)$ in the GCE depends on the commensurability between $L$ and the period of the energetically 
favorable structure.
We shall compare the results obtained in the CE for fixed $N_0$ with the  $p(\rho)$ lines obtained in the GCE for six
system sizes $L$.

\begin{figure}[h]
	\includegraphics[scale = 1]{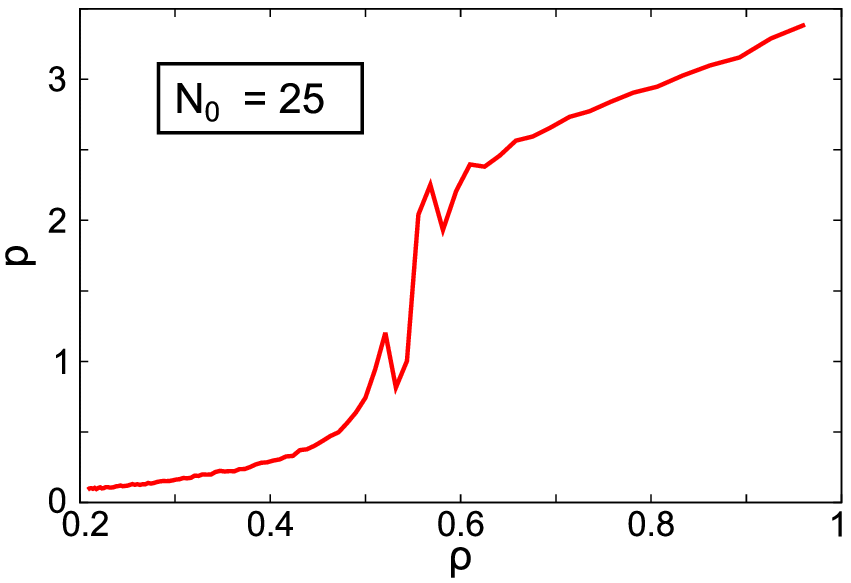}
           \includegraphics[scale = 1]{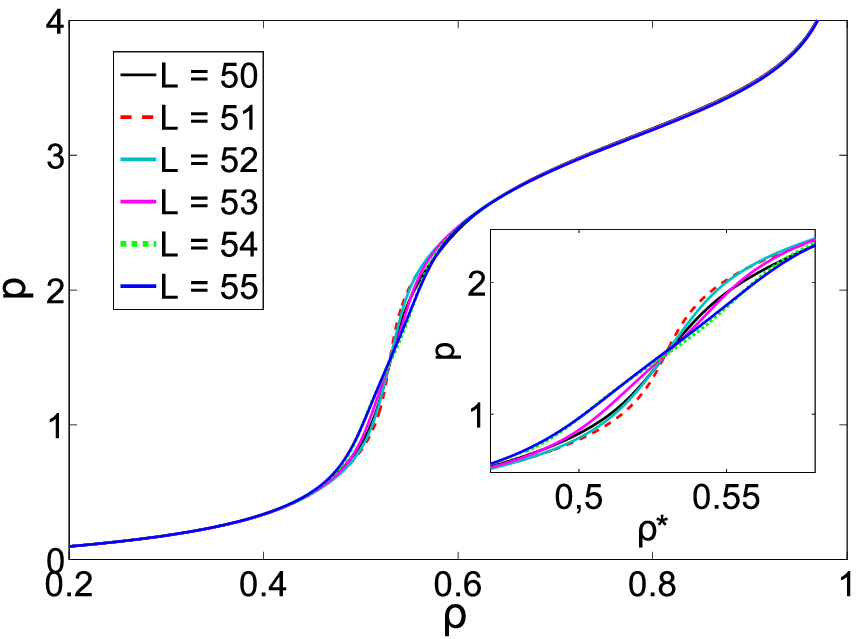}
      \caption{ Reduced pressure as a function of density for $J^*=3$, $T^* =0.5$
      and attractive walls ($h_1^* = h_L^* =- 1$).
      Upper panel: Canonical Monte Carlo (CMC) simulation for the number of particles $N_0 = 25$ 
      and different system sizes $L$, obtained via virtual insertion method. Lower panel: GCE exact results for
      pressure vs. density for different system sizes, $L = 50,51,52,53,54,55$.
 }
	\label{press_den}
  \end{figure}

In Fig. \ref{press_den} the results for  the reduced pressure as a function of the average density
 for CE and GCE are presented. 
Note the discrepancy between the CE and GCE for $\rho \approx 0.55$, where the  periodically ordered 
clusters consisting of 3 particles are separated by 3 empty sites. 
In the GCE $p(\rho)$ increases monotonically, although for $\rho\approx 0.55$ the slope is very large,
and an inflection point is present.
We should stress that in the CE $N_0$ is fixed and the density changes because of the change of $L$.
In contrast, in the GCE $L$ is fixed, and the density changes because $\mu$, and as  a result $\langle N\rangle$,
changes. In the  GCE there are 
different branches of $p(\rho)$ for different $L$. One may interpret the nonmonotonic $p(\rho)$ in the CE as a consequence
of the jumps between the different branches of $p(\rho)$ in the GCE for $L$ and $L-1$. 

\begin{figure}[h]
    \begin{center}
      \includegraphics[scale=0.4]{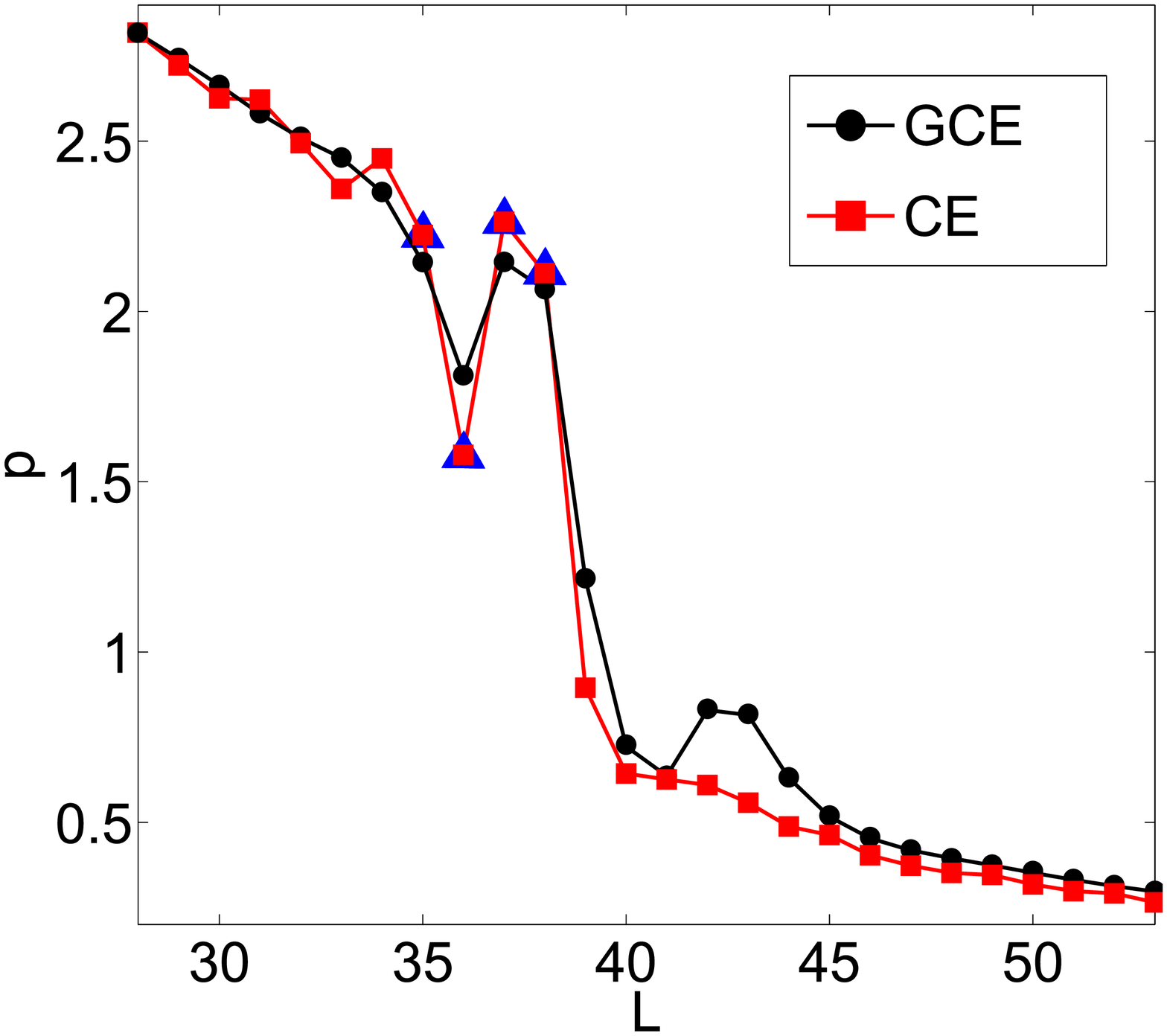},
      \includegraphics[scale=1]{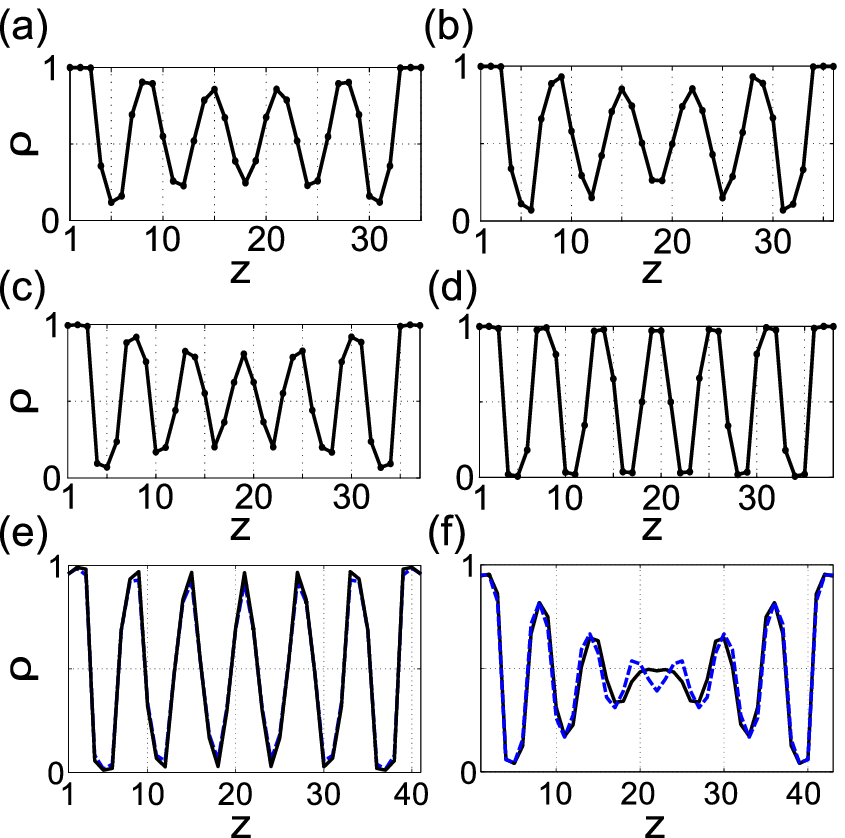}
      \caption{
      Upper panel: reduced pressure as a function of the  system size $L$ for $J^*=3$ and $T^* = 0.5$ 
      in the case of attractive walls ($h_1^* = h_L^* =- 1$). Red curve with squares: Canonical Monte Carlo (CMC)
      simulation for 21 particles. 
      Black curve with bullets: exact results in the GCE with the average number of particles equal to 21. 
      Lower panels: (a-d) CE density profiles  in systems with $L=35$ (a), $L=36$ (b),
      $L=37$ (c) and $L=38$.
      (indicated on the $p(L)$ plot as blue triangles). Panels (e-f): comparison of the density profiles in CE (solid line)
      and GCE (dashed line) for $L=42$ (e) and $L=43$ (f).
}
      \label{w_D_size_dep_aa}
    \end{center}
  \end{figure}

In order to separate the effect of the fluctuation of the number of particles and the effect of the method 
by which the density changes,
we compare the $p(L)$ curves in the CE with $N_0$ particles and in the GCE with $\langle N\rangle =N_0$. 
 In Fig. \ref{w_D_size_dep_aa} the pressure 
is shown as a function of $L$ for the CE with $N_0 =21$ and for GCE with $\langle N\rangle =21$. 
The GCE curve was obtained by finding for each system size $L$ the value of
the chemical potential $\mu_0$ such that $\rho(\mu_0) \approx N_0/L$. We used the exact expression for density 
obtained in Ref.[\onlinecite{pekalski:15:0}]. For such chemical potential the pressure was computed from
the approximate formula $\beta p= -\ln \Xi(\mu_0,L+1,T)+\ln\Xi(\mu_0,L,T)$, which is the 1d lattice version of
the standard expresion $p=-(\partial \Omega/\partial V)_{\mu,T}$.
We also present the density profiles for $L=35,36,37,38$, where $p$ changes rapidly in a nonmonotonic way.
In the case of attractive surfaces the periodic structure where three occupied sites are separated by 
three empty sites is possible for $L=39$, and corresponds to seven clusters. 
For $L<39$ either the clusters are bigger, or the distances between them are smaller. 
In both cases the repulsion between the particles is present, and pressure increases. 
In Fig.\ref{w_D_size_dep_aa} we can see six clusters for $L<37$, and seven 
clusters for $37 \le L\le 42$. 
Note that the nonmonotonic dependence of $p$ on $L$ corresponds to the jump of the number of clusters.
The unusual increase of pressure in the expanding system results from the transition to a larger number 
of smaller clusters. 
The clusters repel each other for $L<39$. 
Upon increase of the system size from $L=38$ to $L=39$ the separation between the clusters becomes
large enough to put the clusters at the separations larger than the range of repulsion, and the pressure drops.

It is interesting that although both the average densities and density profiles for $L = 42$ 
in the two ensembles are the same, the pressure is different. 
The reason is that the pressure depends not on the values of the thermodynamic potentials at a given state,
but on their change and as can be seen on panel (f) of Fig. \ref{w_D_size_dep_aa}, for $L = 43$ the profiles 
differ significantly.

We conclude that the mechanical properties of a confined self-assembling system depend significantly on whether
the system expands for given number of particles, or the separation between the system boundaries is fixed, and the
number of particles decreases due to a change of the chemical potential. 
In both cases we can have the same change of density, but different changes of pressure. 
The unusual increase of pressure upon system expansion is found only in the case of fixed 
(average) number of particles, and is connected with a significant structural reorganization.

\clearpage
 
\section{The case of elastic boundaries}  \label{sec5}
 
\begin{figure}[h]
    \begin{center}
      \includegraphics[scale=1]{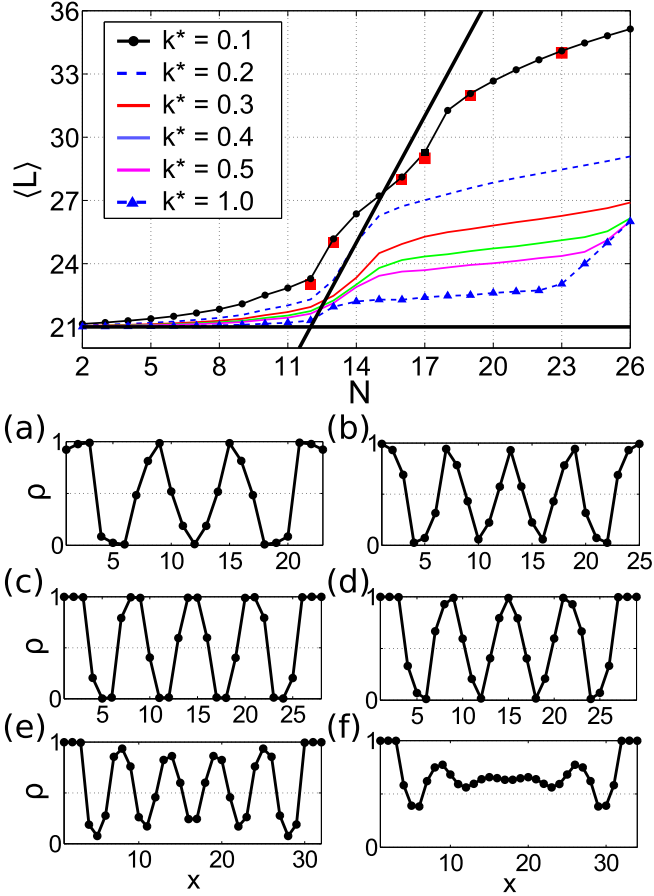}
      \caption{Average system size as a function of the number of particles for different spring constants $k^*$.
      The walls  are attractive, $h^*_L=h^*_1=-1$, $J^*=3$, $T^* = 0.5$ and $L_0 = 21$. 
      Thick solid black lines, $L=L_0$ and $L=2N-3$, correspond to rigid walls and to walls fully 
      adapting to the optimal structure respectively. In the panels (a-f) the density profiles for the number 
      of particls and the system size marked by the red squares along the black solid curve ( $k^*=0.1$) are shown.
      (a) $L=23$ and $N=12$, (b) $N=13$ and $L=25$, (c) $L=28$ and $N=16$, (d) $L=29$ and $N=17$, (e) $L=32$ and $N=19$, 
      (f) $L=34$ and $N=23$.
     }
      \label{osc_size_sc}
    \end{center}
  \end{figure}

In this section we assume that the separation between the system boundaries is not fixed, 
but can fluctuate around $L=L_0$.
The change of the wall separation is associated with the energy cost $\Delta U^*=k^*(L-L_0)^2$. 
Here $k^*$ denotes the elastic constant in units $1/\sigma^2$.
In Fig.\ref{osc_size_sc} the average system size $\langle L(N) \rangle$ as a function of the 
number of particles $N$  is presented.
The confining surfaces are kept at the separation $L$ by the spring that is at rest for $L_0 = 21$. 
We assume attractive walls and $T^*= 0.5$. 
In a system with rigid boundary conditions, attractive walls and $L=21$, 
the periodic structure made of $12$ particles is energetically favorable.  
Thus, for $N \le 12$ only for small values of the spring constant (e.g. $k^*=0.1$) significant
deviations of the average system size from the reference value are present.
For $N > 12$ the internal stress of the fluid  competes with the elastic forces and $\langle L \rangle>L_0$  even 
for $k^*=1$. 

We can distinguish two limiting cases: (i) stiff spring, where the system size saturates 
and the particles become densely packed when $N$ increases, and (ii)
soft spring, where the average system size increases with increasing $N$, and the clusters are separated by empty sites.
If $k^*$ is small enough, then the slope of $\langle L(N) \rangle$ increases when a new cluster made of 3 particles is introduced
to the
system (see panels (a)-(f) of Fig. \ref{osc_size_sc}). On the other hand, for larger values of $k$ the elastic forces are stronger, 
and the system tends to modify the structure of the fluid rather then the system size.
Between the two limiting cases  there is an interesting 
region where the elastic and the solvation force are comparable and compete. 

  \begin{figure}[h]
    \begin{center}
      \includegraphics[scale=1]{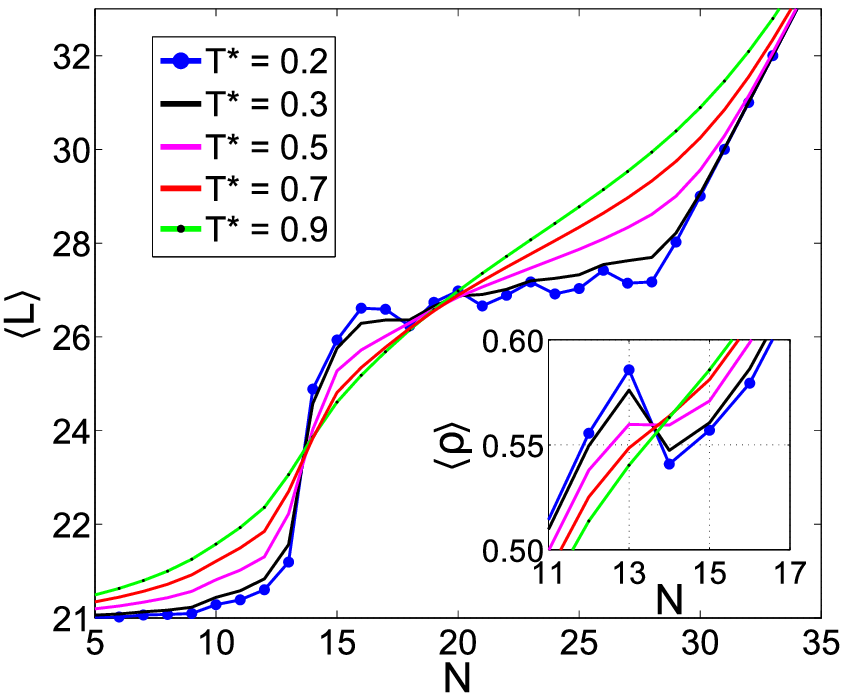}
      \caption{Average system size as a function of the number of particles for different temperatures.
      Attracting walls assumed ($h^*_1=h^*_L=-1$), $J^*=3$, spring constant $k^* = 0.2$ and $L_0 = 21$. 
      The inset presents the average density $\langle \rho \rangle \equiv N/\langle L\rangle$ as a function the number of particles.}
      \label{osc_size_temp}
    \end{center}
  \end{figure}

In Fig.\ref{osc_size_temp} we show  $\langle L(N)\rangle$ for $L_0=21$ and $k^*=0.2$ for several temperatures. At low temperatures three regimes with significantly different slopes of the lines $\langle L(N)\rangle$ can be distinguished.
For $N \lesssim 12 $ the slope of the line $\langle L(N)\rangle$ is small, 
because in this case the separation between the clusters ensures no repulsion between them.
For $ 12 \lesssim N \lesssim 15 $ the system expands significantly 
upon addition of particles, because for $N=14$ an additional cluster appears. In this region the average size of the clusters and the distance between them is 3.
For $N \gtrsim 15$ the slope is small again. Here elastic stress dominates and the clusters average size increases until
the system becomes densely packed.  Note that in this region and at low $T$, before the
system gets filled with particles, we obtain an oscillatory dependence of $\langle L(N)\rangle$ on $N$, with the minima occuring
when the number of particles is a multiple of 3.
Note also that for a given $N$, $\langle L\rangle$  increases with temperature except from $13<N<21$,
where for $L=L_0$  the density
is between the density of the periodic and the closely packed structures. We verified that the anomalous contraction of the heated system is no longer observed at high T.
Finally, note that there is some similarity of
the shapes of the $\langle L(N)\rangle$ and $p(\rho)$ 
lines (Figs.\ref{osc_size_temp} and \ref{press_den}).

\begin{figure}[h]
  \includegraphics[scale=1]{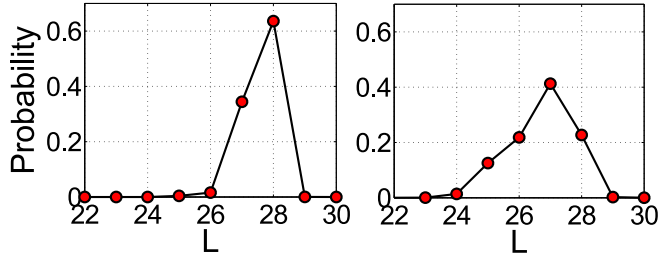}
    \caption{The histograms for the wall separation in the case of elastic boundaries with the spring constant 
  $k^*=0.2$, $N=16$, $L_0=21$, $J^*=3$ and attractive walls ($h^*_1=h^*_L=-1$) for  $T^*=0.2$ (left panel), and $T^*=0.5$ (right panel).}
  \label{his}
\end{figure}

\begin{figure}[h]
  \includegraphics[scale=1]{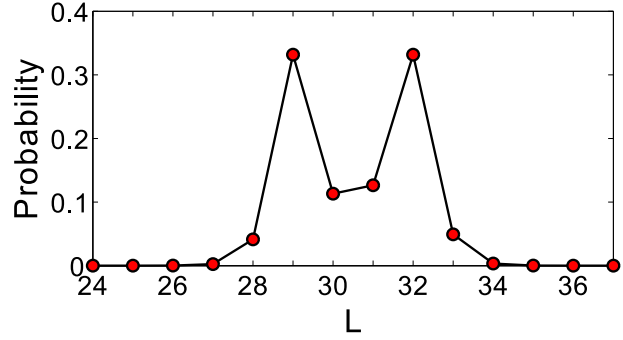}
  \caption{The histogram for the wall separation  in the case of elastic boundaries with the spring constant  $k^*=0.06607$.
   $J^*=3$, $T^*=0.5$, $N=17$, $L_0=19$, attractive walls ($h^*_1=h^*_L=-1$).}
  \label{hisbis}
\end{figure}

The average wall separation and the average density profile give insufficient information about the system behavior.
In Figs.\ref{his} and \ref{hisbis} we present histograms for the wall separation.
Two cases can be distinguished - a single maximum in the probability of the appearance of the 
wall separation $L$, and a bistability with two maxima in this probability, separated by
$\Delta L=3$.


\begin{figure}[h]
\includegraphics[scale=1.2]{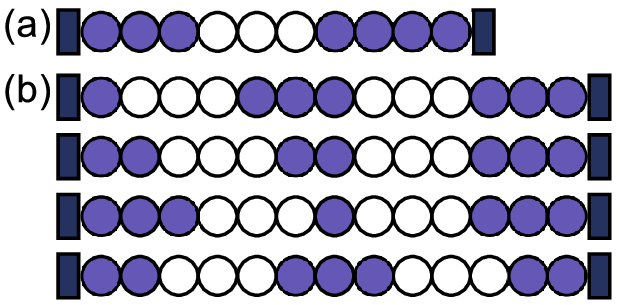}
\includegraphics[scale=1.2]{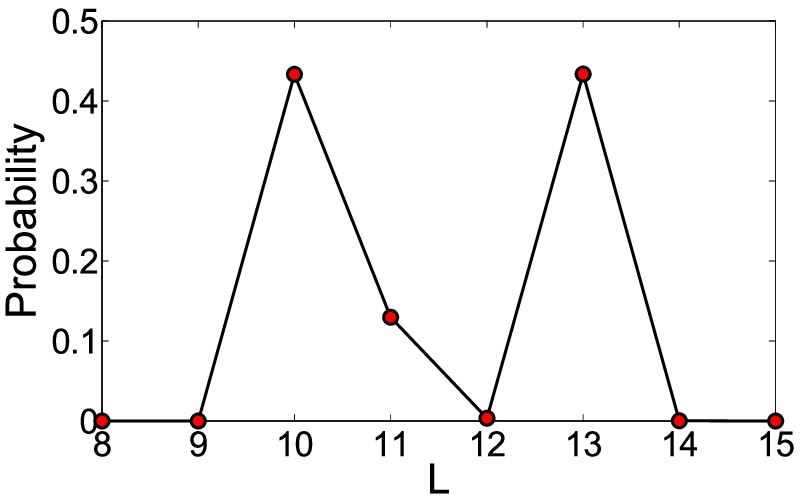}
  \caption{
  Upper panel: 
   Two microstates in the GS of a system consisting of $7$ particles  with the repulsion to attraction ratio  $J^*=3$,
   confined by attractive walls ($h^*_1=h^*_L=-1$) on a spring 
  with the spring constant $k^*=2/9$ that is at rest for $L_0=10$. In panel (a) $L= L_0$ while in panel (b) $L=L_0+3$.
  Note that the microstates symmetric to those also have the same energy, hence
  for  $L=10$ there are 2 different microstates with the same energy, and for  $L=13$ there are 6.
  These microstates correspond to the maxima of the probability shown in the lower panel, in which 
  we present the histogram for the wall separation in the above system in the case of   $T^*=0.2$
  and elastic boundaries with the spring constant  $k^*=2/9+0.0244135$.}
\label{mikro2}
\end{figure}

In order to understand the energetics associated  with the bistability, let us consider the GS for $L_0=10$ and $N=7$. 
The microstates
shown in Fig. \ref{mikro2} correspond to the same  energy of the confined  system (\ref{elasticU}),  when $J^*=3$ and
$k^*=2/9$.  Two different system sizes in the GS
can occur when $N$ is not a multiple of $3$, and for $L=L_0$ an intra-cluster repulsion is present.
The expansion is associated with a simultaneous increase of elastic energy of the walls, and decrease of
the internal energy of the particles,
when the separations between them are such that the repulsion is absent. 
Each microstate  in Fig. \ref{mikro2} occurs with the same probability, but because of
the difference in the degeneracy 
for $L=L_0$ and $L=L_0+3$,
the probability ratio for the two lengths is $p(L_0+3)/p(L_0)=3$.
To estimate the spring constant leading in the above example to equal probability of $L_0$ and $L_0+3$
 for low $T^*$, we  take into account only the microstates shown in Fig. \ref{mikro2}, and require that 
$ \exp(-\beta^*(-5+J^*))=3\exp(-\beta^*(-4+9k^*))$ (see (\ref{elasticU}) for $L=L_0, L_0+3$). For $T^*=0.2$ we obtain 
$k^*\approx 2/9+0.0244136$ in  very good agreement with the results of simulations shown in Fig. \ref{mikro2}. 

\clearpage
\section{Summary and concluding remarks} \label{sec6}

We have studied the effect of various constraints on colloidal self-assembly 
in thermodynamic states that correspond to self-assembly into small clusters separated by voids.
In our model system small clusters with no intra-cluster repulsion yield a negative contribution 
to the internal energy, and do not interact with one another
if the separation between them is larger than the range of repulsion. 
The positions, the size and the number of clusters can fluctuate and different deformations 
of the bulk structure are possible in confinement.
In order to determine the role of constraints imposed on the number of particles and/or the size of the system,
we have compared  density profiles,  equations of state,
and  effects of elastic boundary conditions.

\textbf{Structure:}
\textit{ If the number of particles in the CE is too small for formation of the bulk structure, 
then one more cluster can be present in the GCE despite the same
average number of particles} (Fig \ref{ensemble_compar2a}).  
Interestingly, the different number of clusters in the 
two ensembles is observed for thermodynamic states that do not correspond 
to the maximum of the fluctuation of the number of particles. 
This is because the fluctuations of the number of particles are coupled with the fluctuations of the size of the clusters.
Even a small increase of the number of particles in the GCE together with the splitting of the clusters 
can lead to formation of a larger number of smaller clusters. 
The qualitative difference between the two ensembles occurs for quite large number of particles. 
This behavior is different than in simple fluids,
where the difference between the two ensembles was observed for 
very small number of particles confined in very small pores\cite{gonzales:98:0,kim:99:0}. 

\textbf{Equation of state:}
In the bulk the isotherms $p(\rho)$ do not depend on the way in which the variation of density is attained.
In the confined inhomogeneous systems it is no longer the case:{\textit  Different curves $p(\rho)$ are obtained when 
the size of the system changes
with fixed number of particles, or when the number of particles changes at fixed system size.
 
The shape of the  $p(\rho)$ curve obtained in the GCE with fixed $L$ depends significantly on $L$, or more
precisely on the commensurability of $L$ and the period of the ordered structure (Fig \ref{press_den}). 
In the CE with fixed $N$  we have obtained anomalous decrease of pressure for increasing density for small density intervals 
below and above the density of the 
equilibrium bulk structure (Fig \ref{press_den}).} Inspection of density profiles shows 
that \textit{ the anomalous increase of pressure
for increasing system size with fixed number of particles
is accompanied by increased number of clusters} (the larger cluster splits) (Fig \ref{w_D_size_dep_aa}).
Recall that at short separations the clusters repel each other, and this 
leads to the increase of pressure. The pressure rapidly drops when $L$ further increases and the
clusters do not repel one another any more.
In order to check if the anomalous dependence of pressure on density follows from the fixed number of particles or from
the process by which the density varies, we computed the $p(L)$ curve in the GCE with fixed average number of particles. We have obtained
 similar curves in the two ensembles except for large slits. 
 In both ensembles the anomalous increase of pressure for increasing $L$ is 
 associated with the increasing number of clusters.  For some large $L$ the number of clusters increases with the system size 
 in the GCE but
 not in the CE (Fig.\ref{w_D_size_dep_aa} e,f). In this case a maximum in the $p(L)$ curve is present only in the GCE.

\textbf{Bistability in  elastic confinement:}
If the width of the slit can vary, then the system tends to equilibrate the competing solvation and elastic forces. 
We have found that the equilibrium size of the system is not always unique.
In Ref.\cite{pekalski:15:0} we observed a bistability in a system confined by elastic walls with permeable
walls (fixed $\mu$). 
Two different system sizes can be equally probable: 
one with expanded and the other one with compressed boundaries.
The size fluctuations are accompanied by an absorption or evaporation of a whole cluster.
In the case of impervious walls (fixed $N$) a  bistability exists too. 
\textit{In both cases the origin of the bistability is the change of the 
number of clusters, but the mechanisms which lead to the change are different}. 
The number of clusters can fluctuate for fixed number of particles, when the state with a smaller number 
of bigger clusters and the state with a larger number of smaller clusters are equally probable. 
When the intra-cluster repulsion in large clusters competes with the elastic energy 
of stretched boundaries, the clusters can split and separate. 
The difference between the two equilibrium widths of the system is equal to the
period of the bulk structure in the case of permeable walls, and to half the period of the bulk structure when
the number of particles is constrained.
Since in our 1d system 
the energy barrier is $\sim k_BT$, spontaneous changes of the system size may be induced by thermal fluctuations.

Our results show that\textit{ different anomalies in the confined inhomogeneous 
systems occur when  the release of some constraint or a change of the thermodynamic state leads to 
a change of the number of clusters. 
In particular thermodynamic states the structure and anomalies depend qualitatively on 
the ability of the system to interchange particles with its surroundings.}

Let us stress the difference between the confined simple fluids and the SALR systems. Packing effects of molecules or clusters
play important role in both cases, especially for the solvation force that exhibits oscillatory decay on 
the corresponding length scale in each case. However, the clusters can split or merge, unlike the molecules.
This leads to qualitative differences between the ensembles,  anomalies in the $p(\rho)$  and the $\langle L(N) \rangle$ curves,
and the bistability of the system size that in simple fluids are absent. 

 The patterns emerging in the colloidal and  amphiphilic self-assembly are very similar~\cite{seul:95:0,ciach:13:0,pekalski:14:1}.
 The clusters or layers composed of particles are distributed in space in a similar way as micelles or bilayers composed
 of amphiphilic molecules. Based on the similarity
 between the two types of self-assembly, we can expect that our results may also concern amphiphilic systems such as surfactants or 
 lipids in water and block copolymers, at least on a qualitative level. 
 Similarly, magnetic systems with competing interactions \cite{CaMiStTa2006,CaCaBiSt2011,BaSt2007,BaSt2009}
 may have very similar properties in confinement.
 
\acknowledgments

JP acknowledges the financial support by the National Science Center under Contract Decision No. DEC-2013/09/N/ST3/02551.
N.G.A. acknowledges the support from the Direcci\'on
General de Investigaci\'on Cient\'{\i}fica  y T\'ecnica under Grant
FIS2013-47350-C5-4-R.
AC acknowledges the financial support by the National Science Center grant 2012/05/B/ST3/03302.
JP received funding for the preparation of the doctoral dissertation from the National Science Center in the funding of PhD scholarships
on the basis of the decision number DEC-2014/12/T/ST3/00647.

\appendix 
\section{The expressions for $H^*_j$ and $\sum_{\hat S(M)}'$}
\label{app1}
The $H^*_j$ contains the pair interactions between the particles at the sites of the 
$M$-th box, as well as the interactions between the particles at the sites labeled $3M\!+\!1$ and $3M\!+\!2$
 (if they exist for given $L$).
\begin{displaymath}
H^*_j[\hat S(M)]\!=\! \left\{ \begin{array}{ll}
\!  - ( \sum_{i = 0}^{1} \ro(3M\!-\!i)\ro(3M\!-\!i-\!1))-\mu^* (\sum_{i = 0}^{2} \ro(3M-i)), & \textrm{if $j\! =\! 0$}\\
  -( \sum_{i = 0}^{2} \ro(3M\!+\!1 - i)\ro(3M\!-\!i))+J^*  \ro(3M\!-\!2)\ro(3M\!+\!1) & \textrm{if $j\! =\! 1$} \\
-\mu^* (\sum_{i = 0}^{3} \ro(3M+1-i)),  &\\
  -( \sum_{i = 0}^{3} \ro(3M\!+\!2 - i)\ro(3M\!+1-i))+ & \textrm{if $j \!=\! 2$}\\
J^* (\sum_{i = 0}^{1} \ro(3M\!-\!2+i)\ro(3M\!+\!1+i))-\mu^* (\sum_{i = 0}^{4} \ro(3M+1-i)). &
\end{array} \right.
\end{displaymath}
Whereas $\sum_{\hat S(M)}'$ denotes
\begin{displaymath}
\sum_{\hat S(M)}'= \left\{ \begin{array}{ll}
\sum_{\hat S(M)}  & \textrm{if $j\! =\! 0$}\\
 \sum_{\hat S(M)} \sum_{\ro(3M\!+\!1)}  & \textrm{if $j\! =\! 1$}\\
 \sum_{\hat S(M)} \sum_{\ro(3M\!+\!1)} \sum_{\ro(3M\!+\!2)}  & \textrm{if $j \!=\! 2$}
\end{array} \right.
\end{displaymath}
\section{Microstates obtained by the virtual expansion of the system}
\label{app2}

We consider $N$ indistinguishable particles and  $L$  lattice sites. Each site can be empty or occupied by one particle, 
thus there are ${L \choose N}$  distinguishable microstates.
We will show that the virtual expansion procedure of building configurations of the system with $L+1$ sites 
by inserting an empty site at a random position in a system with $L$ sites is not biased by the insertion procedure.
Let us consider two sets of particle configurations. The elements of the first set are the microstates of a system of
size $L$ with $N$ occupied sites with a distinguished position in which a new site can be inserted. Since the new site 
can be inserted in $L+1$ places, the first set contains $(L+1) {L \choose N}$ different elements. In the second set each 
of the elements corresponds to one of the microstates of a system of $L+1$ sites with $N$ occupied sites, and with one 
of its $L+1-N$ empty sites marked as removable. The number of elements of the second set is $(L+1-N) {L+1 \choose N}$. 
Because  $(L+1) {L \choose N}=(L+1-N) {L+1 \choose N}$, the two sets are equinumerous. 
We define a one-to-one correspondence between the elements
of the two sets by identifying the location of the insertion point in an element of the first set with the location
of the removable site in the element of the second set, and by requiring that the same sites are occupied. Note that 
each microstate of the system with $L+1$ sites
can be obtained in  $L+1-N$ ways from the elements of the second set by removing the mark ``removable''. Because of 
the one-to-one correspondence between the elements of the two sets,
each microstate of the system with $L+1$ sites is obtained $L+1-N$ times by the above procedure.

It follows that the proposed procedures of building configurations of the system with $L+1$ sites 
by inserting an empty site at a random position on the configurations of a system with $L$ sites
are not biased by the insertion procedure. The same lack of bias applies in the reverse procedure.

For an illustration let us consider $L=2$ and $N=1$. 
There are 2 microstates, [1,0] and [0,1]. 
After insertion of an empty site in 3 possible places, we obtain from the first microstate [0,1,0], [1,0,0], [1,0,0], 
and from the second
microstate [0,0,1],[0,0,1], [0,1,0]. One can easily see that after this procedure we obtained 
each microstate in the system of size $L+1=3$ 
containing $N=1$ particle  $L+1-N=2$ times.

\end{document}